\documentclass[12pt]{article}
\usepackage[latin1]{inputenc}
\usepackage{amsmath}
\usepackage{amsfonts}
\usepackage{amssymb,MnSymbol,slashed}
\usepackage{amsxtra}
\usepackage[margin=3cm]{geometry}
\usepackage{color}
\usepackage{hyperref}
\hypersetup{linktocpage=true}
\usepackage{graphicx}
\usepackage{placeins}
\usepackage{upgreek} 
\usepackage{mathrsfs}
\usepackage{accents}
\usepackage{multirow} 
\usepackage{tikz}
\usepackage{braket}
\numberwithin{equation}{section}
\newcommand{\beq}{\begin{equation}}
\newcommand{\eeq}{\end{equation}}
\def\be {\begin{equation}}
\def\ee {\end{equation}}
\def\bs#1\es{\begin{split}#1\end{split}}
\def\ba#1\ea{\begin{align}#1\end{align}}
\def\baed#1\eaed{\begin{aligned}#1\end{aligned}}
\def\bged#1\eged{\begin{gathered}#1\end{gathered}}
\def\bea{\begin{eqnarray}}
\def\eea{\end{eqnarray}}

\def\\nn{\nonumber}

\def\d{\delta}

\def\e{\epsilon}

\def\G{\Gamma}

\def\k{\kappa}

\def\m{\mu}
\def\n{\nu}
\def\o{\omega}

\def\bls{\bigg [}
\def\brs{\bigg ]}
\def\blr{\bigg (}
\def\brr{\bigg )}

\newcommand{\cK}{\mathcal{K}}

\newcommand{\cR}{\mathcal{R}}

\newcommand{\cZ}{\mathcal Z}

\def\cO{{{\mathcal O}}}
\def\cM{\mathcal{M}} 
\def\cN{\mathcal{N}}
\def\cV{\mathcal{V}}

\def\Tr{\text{Tr}}

\def\dim{\text{dim}}

\def\pa{\partial}

\def\fr{\frac}

\def\we{\wedge}

\def\tbzero{{\text{\tiny{(0)}}}}
\def\tbone{{\text{\tiny{(1)}}}}
\def\tbtwo{{\text{\tiny{(2)}}}}

\newcommand{\wh}[1]{ {\hat{#1}}{} }

\let\foo\bar 
\renewcommand{\bar}[1]{ {\foo{  #1} }{} }

\newlength{\dhatheight}

\begin{document}
%%%%%%%%%%%%%%%%%%%%%%%%%%%%%%%%%%%%%%%%%%%%%%%%

\baselineskip=16pt
\setlength{\parskip}{6pt}

\begin{titlepage}
\begin{flushright}
\parbox[t]{1.4in}{
\flushright  	IPMU16-0097}
\end{flushright}

\begin{center}

\vspace*{1.5cm}

{\Large \bf On four-derivative terms in IIB Calabi-Yau orientifold reductions } 

\vskip 1.5cm

\renewcommand{\thefootnote}{}

\begin{center}
 \normalsize 
 Matthias Weissenbacher \textsuperscript{ \!\!\!\ddagger}\footnote{\textsuperscript{\ddagger}matthias.weissenbacher@ipmu.jp}
\end{center}
\vskip 0.5cm

Kavli Institute for the Physics and Mathematics of the Universe, University of Tokyo,
Kashiwa-no-ha 5-1-5, 277-8583, Japan
\end{center}

\vskip 1.5cm
%\addtocounter{footnote}{-1}
\renewcommand{\thefootnote}{\arabic{footnote}}

\begin{center} {\bf ABSTRACT } \end{center}
We perform a Kaluza-Klein reduction of IIB supergravity including purely gravitational $\alpha '^3$-corrections on a Calabi-Yau threefold, and perform the orientifold projection accounting for the presence of $O3/O7$-planes. We consider infinitesimal K\"{a}hler deformations of the Calabi-Yau background and derive the complete set of four-derivative couplings quadratic in these fluctuations coupled to gravity. In particular, we find  four-derivative couplings of the K\"{a}hler moduli fields in the four-dimensional effective supergravity theory, which are referred to as friction couplings in the context of inflation.

\end{titlepage}

\newpage
\noindent\rule{\textwidth}{.1pt}		
\tableofcontents
\vspace{20pt}
\noindent\rule{\textwidth}{.1pt}

\setcounter{page}{1}
\setlength{\parskip}{9pt} 

%%%%%%%%%%%%%%%%%%%%%%%%%%%%%%%%%%%%%%%%%%%%%%%%
\section{Introduction}
%%%%%%%%%%%%%%%%%%%%%%%%%%%%%%%%%%%%%%%%%%%%%%%%

The dimensional reduction of ten-dimensional IIB supergravity on Calabi-Yau orientifolds  yields four-dimensional $\cN=1$ supergravity theories \cite{Grimm:2004uq}, which are of particular phenomenological interest.  The resulting couplings  are given by topological quantities of the internal space which are computable for explicit backgrounds, and thus  provide a fruitful environment for string model building \cite{Blumenhagen:2005mu,Grana:2005jc,Douglas:2006es,Blumenhagen:2006ci,Akerblom:2007np,McAllister:2008hb}.  The compactification on a Calabi-Yau threefold preserves a quarter of the supersymmetry of ten dimensions and thus results in a  $\cN=2$ theory in four dimensions, which is then broken to $\cN=1$ by the presence of orientifold planes. Incorporating gauge fields by adding D-branes in the Calabi-Yau background one is led to introduce extended objects with negative tension to cancel gravitational and electro/magnetic tadpoles, given by the orientifold planes, which however carry no physical degrees of freedom by  themselves \cite{Giddings:2001yu}. String theory provides an infinite series in $\alpha'$ of higher-derivative corrections to the leading order two-derivative IIB supergravity action. However, even the next to leading order $\alpha'^3$-correction to the  four-dimensional action arising in Calabi-Yau (orientifold) compactifications are only marginally understood, but have proven to be of high relevance to string phenomenology \cite{Denef:2004ze,Balasubramanian:2005zx}.

In this work, we discuss  a set of four-derivative couplings that arise in four-dimensional $\cN=2$ and $\cN=1$ supergravity theories resulting from purely gravitational eight-derivative  $\alpha'^3$-corrections to ten-dimensional IIB supergravity \cite{Kiritsis:1997em,Antoniadis:1997eg,Liu:2013dna}, upon compactification on a Calabi-Yau threefold and  orientifold, respectively. 
Such corrections are of conceptual  as well as of phenomenological importance. Four-dimensional $\cN=1$ and $\cN=2$ supergravity theories with four-derivative interaction terms are only marginally understood \cite{Katmadas:2013mma,Koehn:2012ar,Horndeski:1974wa,Alvarez-Gaume:2015rwa}, and the knowledge of the relevant couplings is desirable. A recent progress is the classification of  $4d,\,\cN=1$ four-derivative superspace operators for ungauged chiral multiplets \cite{Ciupke:2016agp}. On the other hand higher-derivative couplings have a prominent role in phenomenological models such as inflation \cite{Amendola:1993uh,Bielleman:2016grv, Aoki:2015eba, Dalianis:2014sqa} and  have been used in the context of moduli stabilization recently \cite{Ciupke:2015msa}. 

Dimensionally reducing  ten-dimensional IIB supergravity on a supersymmetric background must yield an effective four-dimensional $\cN=1$, $\cN=2$ supergravity theory depending on how much supersymmetry is preserved by the background. However, the supersymmetric completion at order $\alpha'^3$ of IIB supergravity is not known, thus a exhaustive study of the four-derivative effective action at order $\alpha'^3$  in four dimensions is out of reach.
Hence our strategy will be to focus on a complete subset of the ten-dimensional IIB supergravity theory at order $\alpha'^3$ and argue that the resulting couplings in the four-dimensional theory cannot be altered by any other sector of the higher-dimensional theory.
More concretely, the terms we analyze in ten dimensions carry four Riemann tensors, thus are schematically of the form $\cR^4$ and are shown to be complete \cite{Gross:1986iv,Rajaraman:2005up,Hyakutake:2006aq}. In other words all other possible $\cR^4$-terms are related to this sector via a higher-derivative field-redefinition of the metric.  We hence restrict our analysis to a subset of four-dimensional couplings, which can only origin from the $\cR^4$ -sector and thus must also be complete in the above sense. In particular we  focus on K\"{a}hler deformations of the internal space, which give rise to a set of real scalar fields in the external space. We do not allow for background fluxes or localized sources for D-branes in this work, furthermore we neglect higher-derivative corrections arising due to D-branes and O-planes.

It is well known that the classical Einstein-Hilbert term gives rise to the kinetic terms for the K\"{a}hler moduli. The $\cR^4$-sector generically corrects the couplings of the kinetic terms at order $\alpha'^3$  by some expression carrying  six internal space derivatives \cite{Bonetti:2016dqh}, which was also discussed in the context of M-theory/F-theory in \cite{Grimm:2013bha,Grimm:2013gma,Grimm:2014efa,Grimm:2014xva,Grimm:2015mua}. However, these $\alpha'$-corrections will not be addressed  in this work. Furthermore, note that the two-derivative kinetic terms  generically receive backreaction effects at order $\alpha'^3$ from the modified supersymmetric background at this order in the string length. However, the four-derivative external terms arising from $\cR^4$ do not receive corrections from the modified background since these would be even higher order in $\alpha'$. The interaction terms of the  K\"{a}hlermoduli fields with the four-dimensional metric moreover can only arise form purely gravitational terms in ten dimensions given at order $\alpha'^3$ solely by  the $\cR^4$-sector. We restrict ourselves to study four-derivative couplings at most quadratic in the infinitesimal K\"{a}hlermoduli deformations. However, a complete analysis would need to also take into account cubic and quartic infinitesimal K\"{a}hlermoduli deformations, which will be discussed in a forthcoming work \cite{MCY3}.

This paper is organized as follows. In section \ref{4dLagr} we review the relevant  $\cR^4$-terms in ten dimensions, comment on the supersymmetric background, and  discuss the four-derivative couplings quadratic in the K\"{a}hlermoduli deformations, arising upon  dimensional  reduction on a Calabi-Yau threefold. In section \ref{4dN1} we then perform the orientifold projection to yield the $\cN=1$ couplings at fourth order in derivatives.

%%%%%%%%%%%%%%%%%%%%%%%%%%%%%%%%%%%%%%%%%%%%%%%%
\section{The 4d four-derivative Lagrangian}\label{4dLagr}
%%%%%%%%%%%%%%%%%%%%%%%%%%%%%%%%%%%%%%%%%%%%%%%%

 This section discusses the dimensional reduction of IIB supergravity including purely gravitational eight-derivative corrections on a Calabi-Yau threefold to four dimensions. We fluctuate the background metric by K\"{a}hler deformations and focus on couplings which carry four external space derivatives and are at most quadratic in the infinitesimal K\"{a}hler deformations.
We first review the relevant $\alpha'^3$ $R^4$-corrections to ten-dimensional IIB supergravity and the supersymmetric background. 
%%%%%%%%%%%%%%%%%%%%%%%%%%%%%%%%%%%%%%%%%%%%%%%%
\subsection{ IIB higher-derivative action}\label{redres}
%%%%%%%%%%%%%%%%%%%%%%%%%%%%%%%%%%%%%%%%%%%%%%%%

The IIB higher-derivative action at order $\alpha '^3$ has various contributions \cite{Schwarz:1982jn,Grisaru:1986kw,Gross:1986mw,Abe:1987ud,Kehagias:1997jg,Kehagias:1997cq,Minasian:2015bxa,Policastro:2006vt,Policastro:2008hg}. For the discussion at hand only the $\cR^4$-sector containing four ten-dimensional Riemann tensors will be relevant.  This subsector of the IIB supergravity action at order $\alpha'^3$ in the Einstein-frame is given by
\beq\label{RR4action}
S_{\text{grav}} = S_{EH} +  \alpha \;  S_{\wh R^4} \;\; , \;\ \; \text{with} \;\;\;  \alpha =  \frac{\zeta(3) \alpha'^3}{3 \cdot 2^{10}} \;\; ,
\eeq
and 
\beq
S_{EH} =  \fr{1}{2 \k_{10}^2}  \int  \wh R \wh\ast 1 \;\; ,
\eeq
where $2\kappa_{10}^2 = (2\pi)^7 \alpha'^4$.  The higher-derivative contribution can be schematically written as
\beq\label{R4}
S_{\wh R^4} =  \fr{1}{2 \k_{10}^2}  \int e^{-\frac{3}{2} \wh\phi} 
\big( t_8  t_8 + \tfrac{1}{8}  \e_{10}  \e_{10} \big) \wh R^4 \wh * 1  \;\; ,
\eeq
where the explicit tensor contractions are given by
\ba\label{deft8t8R4}
 \e_{10} \e_{10} \wh R^4 &=  \epsilon^{R_1 R_2   M_1\ldots M_{8} }  \epsilon_{R_1 R_2 N_1 \ldots N_{8}} \wh R^{N_1 N_2}{}_{M_1 M_2} \wh R^{N_3 N_4}{}_{M_3 M_4}  \wh R^{N_5 N_6}{}_{M_5 M_6} \wh R^{N_7 N_8}{}_{M_7 M_8} \ , \nonumber \\[.1cm]
  t_8  t_8 \wh R^4 & =  t_{8}^{  M_1 \dots M_8}  t_{8 \, N_1  \dots N_8}     \wh R^{N_1 N_2}{}_{M_1 M_2} \wh R^{N_3 N_4}{}_{M_3 M_4}  \wh R^{N_5 N_6}{}_{M_5 M_6} \wh R^{N_7 N_8}{}_{M_7 M_8} \ .
\ea
Where $  \e_{10}$ is the ten-dimensional Levi-Civita tensor and the explicit definition of the tensor $t_8$ can be found in \cite{Freeman:1986zh}.
Let us note that we do not discuss higher-derivative terms of the dilaton, since we lack  completeness of the ten-dimensional action. However, the complete axio-dilaton dependence of the $R^4$-terms is known to be
\beq\label{R42}
S^\tbtwo_{\wh R^4} =  \fr{1}{2 \k_{10}^2}\int  E(\tau,\bar \tau)^{3/2}  
\Big( t_8  t_8 + \tfrac{1}{8}  \e_{10}  \e_{10} \Big) \wh R^4 \wh * 1 \;\; ,
\eeq
where $E(\tau,\bar \tau)^{3/2}$ is the $SL(2,\mathbb Z)$-invariant  Eisenstein Series   given by 
\beq\label{Eisen32}
E(\tau,\bar \tau)^{3/2} =
 \sum_{(m,n) \neq(0,0)} \frac{\tau_2^{3/2}}{|m + n \, \tau|^3} \ ,
\eeq
with  $\tau = \wh C_0 + i e^{ -  \wh \phi} := \tau_1 + i \tau_2$ the axio-dilaton.
In the large $\tau_2$ limit, which corresponds to the small
string coupling limit \eqref{Eisen32} results in 
\beq \label{f0expansion}
E(\tau,\bar \tau)^{3/2} = 2 \zeta(3) \, \tau_2^{3/2} + \tfrac{2\pi^2}{3} \tau_2^{-1/2}
+ \cO(e^{-2\pi \tau_2}) \ .
\eeq 
We will use this approximation  in \eqref{R42} in the following discussion, and only look at the leading order contribution in $g_s$, the string coupling, given by  \eqref{R4}. 

%%%%%%%%%%%%%%%%%%%%%%%%%%%%%%%%%%%%%%%%%%%%%%%%
\subsection{Supersymmetric background}\label{redres}
%%%%%%%%%%%%%%%%%%%%%%%%%%%%%%%%%%%%%%%%%%%%%%%%

The supersymmetric background of  ten-dimensional  IIB supergravity  at the two-derivative level, thus at leading order in $\alpha'$ is given by a Calabi-Yau  threefold $Y_3$. For simplicity we do not consider localized sources and background fluxes, and thus the line element is given by
\ba\label{metricbackg}
ds^2 =  \eta_{\m \n} dx^\m dx^\n +  2 g^\tbzero_{m \bar n} dy^m dy^\bar n \;\; ,
\ea
with  $\eta_{\mu \nu}$  the Minkowski metric, where $\m = 0,1,2,3$  is a $4d$ external space world index and $m = 1,\ldots, 3$ is the index of the complex  three-dimensional internal Calabi-Yau manifold with metric $g^\tbzero_{m \bar n} $, where  $m,\bar m$ are holomorphic and anti-holomorphic indices, respectively. Taking into account the higher-curvature corrections \eqref{R4} in ten dimensions,  \eqref{metricbackg}  is no longer a supersymmetric background but needs to be modified such that the internal manifold  is no longer  Ricci flat.
It was shown that the internal space metric is modified as
\beq
g^\tbzero_{m \bar n} \longrightarrow g^\tbzero_{m \bar n} + \alpha'^3 g^\tbone_{m \bar n} \;\; ,
\eeq
where $g^\tbone_{m \bar n}$ is a solution to the modified Einstein equation $R_{m \bar n} =   \alpha'^3\partial_m \partial_{\bar n}  Q$, with Q the six-dimensional Euler-density \eqref{Euler},  \cite{Freeman:1986br}.
 However, we can restrict our analysis to the case of the leading order metric \eqref{metricbackg}, since at order $\alpha'^3$ the four-derivative couplings only receive corrections from the $\cR^4$-terms evaluated on the zeroth order Calabi-Yau background.
We do not  incorporate for  internal flux in this work, since the considered sector decouples, and also do not allow for localized D-brane sources, which would give rise to a warpfactor in \eqref{metricbackg}.

 In the following we freeze the complex structure moduli, and allow solely for  the K\"{a}hler deformations given by the harmonic $(1,1)$-forms $\{\omega_{i}\}$, with  $i=1,...,h^{1,1} $, where $h^{(1,1)} = \dim H^{(1,1)}$ the dimension of the $(1,1)$-cohomology group. The harmonicity is w.r.t.\,the zeroth order Calabi-Yau metric. These give rise to the massless K\"{a}hler moduli fields by varying the background metric by
\beq\label{Kaehlerfluc}
g^\tbzero_{m \bar n } \rightarrow g^\tbzero_{m\bar n} - i  \, \d v^i \, \omega_{i \, m \bar n}\;\;,
\eeq
 where $\d v^i$ are the real scalar infinitesimal K\"{a}hler deformations.\footnote{Note that we choose the fluctuation to be $- i  \, \d v^i \, \omega_{i \, m \bar n}$. The choice of sign is such that combined  with the convention $J_{m \bar n} = i g_{m \bar n}$, to give a positive sign in $\d J = \d v^i \o_i$.} Let us emphasize that also  \eqref{Kaehlerfluc} receives $\alpha'^3$-corrections \cite{Grimm:2014efa}, however, these do not affect  the four-derivative couplings at the relevant order in $\alpha'$. A preliminary study for allowing both the complex structure deformations and K\"{a}hler deformations simultaneously at the higher-derivative level arising from the $\cR ^4$-sector in the context of M-theory can be found in \cite{mythesis}.
In this work we consider four-derivative couplings which are up to quadratic order in the infinitesimal K\"{a}hler deformations $\d v^i$.

%%%%%%%%%%%%%%%%%%%%%%%%%%%%%%%%%%%%%%%%%%%%%%%%
\subsection{Reduction results}\label{redres}
%%%%%%%%%%%%%%%%%%%%%%%%%%%%%%%%%%%%%%%%%%%%%%%%

Compactifying the action \eqref{RR4action} on the Calabi-Yau background \eqref{metricbackg} we expand the result at four external derivative level up to quadratic order in the infinitesimal K\"{a}hler deformations \eqref{Kaehlerfluc}.
The reduction result may be expressed entirely in terms of the second Chern-form $c_2$, see  \eqref{Chernclasses},  the K\"{a}hler form \eqref{eq:Kform} and a higher-derivative object $Z_{m \bar m n \bar n}$ \cite{Katmadas:2013mma}
given by 
\beq\label{higherder}
Z_{m \bar m n \bar n} =  \tfrac{1}{(2 \pi)^2}\e _{ m \bar m m_1 \bar m_1 m_2 \bar m_2 }  \e _{ n \bar n n_1 \bar n_1 n_2 \bar n_2 } R {}^{\bar m_1 m_1 \bar n_1 n_1} R {}^{\bar m_2 m_2 \bar n_2 n_2}\; .
\eeq
Its analog for a Calabi-Yau four-fold has been encountered in the context of M-theory/F-theory in \cite{Grimm:2014efa}. $Z_{m \bar m n \bar n}$  in \eqref{higherder} obeys the following relations
\ba\label{Zrel}
&Z_{m \bar m n \bar n}= -Z_{m \bar n  n \bar m} = Z_{n \bar m m \bar n} &\;\;\;\;\; &Z_{m \bar m} =Z_{m \bar m n}{}^{ n} = -2 i (\ast c_2)_{m \bar m}& \;\; & Z_{m\bar m} \omega_i^{\bar m m} = 2 i\ast (c_2\wedge \omega_i)&  \nonumber \\ \nonumber \\ 
& Z_{m \bar m} g^{\bar m m}= Z_{m}{}^m{}_n {}^{ n} = 2 \ast(c_2 \wedge J) & \;\;\; & Z_{m \bar m n \bar n}R^{m \bar m n \bar n} =  - 3!  \, 2 \pi \ast c_3  \;\; .
\ea
Note that $Z_{m \bar m n \bar n}$  has the symmetry properties of the Riemann tensor build from a K\"{a}hler metric. It is itself not topological but is related to second and third Chern form of a Calabi-Yau  manifold of dimension $n \geq 3$. In the following we dress objects constructed from the background Calabi-Yau metric with the symbol - $ ^\tbzero$  - as e.g. $Z^\tbzero_{m \bar m n \bar n}$.

We have now set the stage to discuss the reduction results.
By fluctuating the Calabi-Yau metric with the K\"ahler deformations, the higher-derivative $\alpha'^3$-terms \eqref{R4} at two-derivative level  give rise to a $ \alpha'^3 $-modified  four-dimensional Einstein-Hilbert term \cite{Becker:2002nn} and $ \alpha'^3 $-corrections to the kinetic terms for the K\"{a}hler moduli fields \cite{Bonetti:2016dqh}.  The explicit form of these corrections has been  also worked out in the context of M-theory on Calabi-Yau fourfolds in \cite{Grimm:2013bha,Grimm:2013gma,Grimm:2014efa,Grimm:2015mua}. The four-dimensional dilaton $\phi$ arises as $\wh\phi \to\phi$. Its internal component is constant at leading order but is given by $\phi \propto  \alpha'^3 Q $ at the order of consideration. However, for the discussion at hand only the leading order constant part is relevant.  The focus of this work is to derive the four-derivative corrections to the leading order two-derivative 4d Lagrangian, as discussed next.
The reduction of the classical Einstein-Hilbert term gives
\beq\label{classR}
 \fr{1}{2 \k_{10}^2}  \int  \wh R \wh\ast 1 \longrightarrow \fr{1}{2 \k_{10}} \int_{\cM_4}  \Big[ \Omega  R  +  \nabla_\mu \d v^i  \nabla^\mu \d v^j \int_{Y_3}   \Big( \tfrac{1}{2}  \o_{i m \bar n}  \o_{j }{}^{\bar n m}  - \o_{i m}{}^m  \o_{j n}{}^n  \Big) \Big] \ast_4 1\;\; + \cO (\alpha) ,
\eeq
with
\beq\label{Weyl0n}
  \Omega  = \int_{Y_3} \Big[ 1  - i   \d v^i  \, \o _{i m}{}^{m}  +  \tfrac{1}{2}  \d v^i \d v^j  ( \o _{i m \bar n}  \o _{j }{}^{\bar n m} -  \o _{i m}{}^m  \o _{j n}{}^n ) \Big] *_6  1\ ,
\eeq
where the $\cO(\alpha)$ corrections in  \eqref{classR} arise due to the mentioned $\alpha'^3$-modification of the background. However, these terms do not interfere with our analysis. It is necessary to consider the Weyl rescaling factor \eqref{Weyl0n} up to order $ (\d v)^2 $.
The four-derivative corrections arising from the ten-dimensional $\cR^4$-terms result in
\ba\label{R4Red}
& \tfrac{1}{2 \k_{10}^2}  \int e^{-\frac{3}{2} \wh\phi} 
 \Big( t_8  t_8 + \tfrac{1}{8}  \e_{10}  \e_{10} \Big) \wh R^4 \wh * 1 \quad \quad \quad  \longrightarrow   
 \ea\vspace{- 0,5 cm}
 \ba
 \tfrac{192 (2 \pi)^2}{2 \k_{10}^2} \int_{M_4}   e^{-\frac{3}{2} \phi}  \bls \, \;\;  &\Big[  4  \, R_{\mu \nu} R^{\mu \nu} - R^2  \Big]\Big( \int_{Y_3} c_2^\tbzero\wedge J ^\tbzero + \d v^i   \int_{Y_3} c_2^\tbzero\wedge \o_i  + \d v^i  \d v^j \int_{Y_3}  \delta_j (c_2^\tbzero\wedge \o_i  )  \Big) &  \nonumber \\ 
%+   R_{\mu \nu \rho \sigma} R^{\mu \nu  \rho \sigma} \int_{Y_3} c_2\wedge J +
\;\; + &\Big[   \big( - 2 R_{\mu \nu }   + \tfrac{1}{2} R g_{\mu \nu} \big)\nabla^\mu \d v^i  \nabla^\nu \d v^j  \ + \nabla_\mu \nabla^\mu \d v^i  \; \nabla_\nu \nabla^\nu \d v^j \ \Big]\, \int_{Y_3}   \, Z^\tbzero_{m\bar m n \bar n} \o_i{}^{\bar m m} \o_j{}^{\bar n n} \ast  1 & \nonumber \\
& \;\; -2  \nabla_\mu \nabla_\nu \d v^i  \; \nabla^\mu \nabla^\nu \d v^j \ \, \int_{Y_3}   \, Z^\tbzero_{m\bar m n \bar n} \o_i{}^{\bar m m} \o_j{}^{\bar n n} \ast  1  \;\;\; \brs\ast_4  1 \;\; ,& \nonumber
\ea
 where $\d_i$ denotes the variation resulting from the metric shift \eqref{Kaehlerfluc}. Note that 
 \beq
 \int_{Y_3}  \delta_j (c_2^\tbzero \wedge \o_i  )  =0 \;\; ,
 \eeq
  since $c_2\wedge \o_i  $ is a topological quantity and hence its variation results in a total derivative. Furthermore, let us note that the four-dimensional Euler-density  is given by 
\beq\label{euler}
 e(\nabla) = R^2  -4 R_{\mu \nu} R^{\mu \nu}  + R_{\mu \nu \rho \sigma } R^{\mu \nu \rho \sigma} \; , \;\; \text{with} \;\;  \int_{M_4} e(\nabla) \ast_4 1= \chi(M_4) \;\; ,
\eeq
where $\chi(M_4)$ is the Euler-characteristic of the external space $M_4$. Comparing \eqref{euler} to \eqref{R4Red} one infers that one may express the reduction result at zeroth order in $\d v^i$ in terms of $R_{\mu \nu \rho \sigma } R^{\mu \nu \rho \sigma} $,  plus the topological term dependent on $\chi(M_4)$. However, we will not perform this substitution since there is a more intuitive way of expressing the result as we will discuss in the next section.
Let us stress that \eqref{R4Red} is not the complete reduction result at the four-derivative level arising from the $\cR^4$-sector, but we have neglected terms cubic and quartic in the fluctuations $\d v^i$. Their derivation is crucial for a complete understanding, and we refer the reader to future work.

\subsubsection{Weyl rescaling }

In this section we perform the Weyl rescaling of the four-dimensional action composed of \eqref{classR} and \eqref{R4Red} to the canonical Einstein-frame. Furthermore, we discuss the extension of the infinitesimal K\"{a}hler  deformations to finite fields.
The Weyl rescaling of the classical Einstein-Hilbert term gives
\beq\label{Weyl2}
\tfrac{1}{2 \k_{10}^2} \int_{\cM_4}  \Omega  R *1    \stackrel{\text{Weyl}}{\rightarrow} \tfrac{1}{ (2 \pi)^4 \alpha'} \int_{\cM_4}   R *1   - \tfrac{3  }{2} \nabla_\mu \d v^i  \nabla^\mu \d v^j  \;  \tfrac{1}{ \cV^\tbzero{}^2} \cK^\tbzero_{i} \cK^\tbzero_{j} *1 \;\; .
\eeq

Where we have used identities \eqref{IN2} for the intersection numbers $\cK^\tbzero_i, \, \cK^\tbzero_{ij}, \,   \cK^\tbzero_{ijk}$, whose definitions are given in \eqref{IN1}. Moreover, note that  from  the definition \eqref{IN1} it is manifest that the volume $\cV^\tbzero$ and the intersection numbers $\cK^\tbzero_i,\cK^\tbzero_{ij},\cK^\tbzero_{ijk}$ are dimensionless and are expressed in terms of the length scale $\alpha'$. In this conventions also the fields $\d v^i$ are dimensionless.

Due to the appearance of the four-derivative term the Weyl rescaling of the action is more involved. One may show that  by using \eqref{Weyl1} and \eqref{Weyl2} up to total derivative contributions at order  $\alpha'^3$ one finds
\bea\label{Weyln}
\int_{M_4}  e^{-\frac{3}{2} \phi} \bls & \Big[  4  \, R_{\mu \nu} R^{\mu \nu}- R^2   \Big]  \Big( \int_{Y_3} c^\tbzero_2\wedge J^\tbzero  + \d v^i   \int_{Y_3} c^\tbzero_2\wedge \o_i  \Big)  \quad \quad \brs  \ast 1\quad   \quad \quad \quad&  \\
& \stackrel{\text{Weyl}}{\rightarrow}&  \nonumber \\
\int_{M_4} e^{-\frac{3}{2} \phi} \bls & + \;\Big[  4  \, R_{\mu \nu} R^{\mu \nu} - R^2  \Big] \;\; \Big( \int_{Y_3} c^\tbzero_2\wedge J^\tbzero  + \d v^i   \int_{Y_3} c^\tbzero_2\wedge \o_i  \Big) \quad \quad \quad \quad \quad \quad  \quad \quad &\nonumber \\
& -  \;\; \Big[4R_{\mu \nu}   - R g_{\mu \nu} \Big] \nabla^\mu \nabla^\nu \d v^i \;\;   \frac{2}{\cV^\tbzero}  \cK^\tbzero_{i} \Big( \int_{Y_3} c^\tbzero_2\wedge J^\tbzero  + \d v^i   \int_{Y_3} c^\tbzero_2\wedge \o_i  \Big) \quad \quad  &\nonumber \\
& \,- \;\; \Big[4R_{\mu \nu}   - R g_{\mu \nu} \Big] \nabla^\mu \d v^i  \nabla^\nu \d v^j  \; \; \frac{1}{\cV^\tbzero{}^2}  \cK^\tbzero_{j}  \cK^\tbzero_{i} \,\int_{Y_3} c^\tbzero_2\wedge J^\tbzero \quad \;\;  \quad \quad \quad  \quad\;\;\; \;\;& \nonumber \\
& \;\;+ \;\;  \Big[ 4  \nabla_\mu\nabla_\nu \d v^i \;\;  \nabla^\nu \nabla^\mu \d v^j - \nabla_\mu\nabla^\mu \d v^i \;\;  \nabla_\nu \nabla^\nu \d v^j    \Big]  \;\;  \frac{1}{\cV^\tbzero{}^2}  \cK^\tbzero_{i} \cK^\tbzero_{j}   \int_{Y_3} c^\tbzero_2\wedge J^\tbzero   &  \brs\ast_4  1 + \dots \nonumber
\eea
 The elipses denote terms where more than two fields $\d v^i$ carry derivatives and furthermore terms, which have derivatives acting on the dilaton. An exhaustive derivation of the four-derivative dilaton action would require the knowledge of the ten-dimensional higher-derivative dilaton action \cite{Gross:1986mw,Kehagias:1997jg,Kehagias:1997cq,Minasian:2015bxa,Policastro:2006vt,Policastro:2008hg}, which lacks completeness and is  hence beyond the scope of our study. %Using \eqref{Weyl1} and \eqref{Weyl2} one derives the action in the canonical Eintein-Hilbert form for gravity and the Kaehlermoduli sector...

Before collecting the contributions arising due to the Weyl rescaling  \eqref{Weyl1}, \eqref{Weyl2} and  combining it with the reduction results  \eqref{classR} and \eqref{R4Red}   let us first lift the infinitesimal K\"ahler fluctuations around the background metric to full fields. We proceed by making the naive replacement $v^i = v^\tbzero{}^i + \d v^i $, where $ J^\tbzero = v^\tbzero{} ^i \o_{i}$ is the background K\"{a}hler form. 
This substitution is straightforward when the couplings are given by topological quantities as in the case of them being intersection numbers, where one simply infers e.g. $\cK^\tbzero_i \to \cK_i$.
Analogously, one infers in the case of the topological higher-derivative coupling that
\beq
 \int_{Y_3} c^\tbzero _2\wedge J^\tbzero  + \d v^i   \int_{Y_3} c^\tbzero _2\wedge \o_i  \longrightarrow  \int_{Y_3} c_2\wedge J \;\; ,
 \eeq
where $ J = v^i \o_{i}$, and $c_2$ is constructed from the metric $g_{m\bar n} = - i v^i \o_i$. However, the uplift of the coupling $\int_{Y_3}  \, Z^\tbzero_{m\bar m n \bar n} \o_i{}^{\bar m m} \o_j{}^{\bar n n}$ is less trivial since it does not represent a topological quantity of the internal Calabi-Yau threefold. 
We will write the uplift of this coupling in the action by naively replacing the background metric by $g_{m\bar n}$, thus one yields $\int_{Y_3}  \, Z_{m\bar m n \bar n} \o_i{}^{\bar m m} \o_j{}^{\bar n n}$. However, a more refined analysis would be required to fully justify this choice.

Combining the uplift of the reduction result   \eqref{classR}, \eqref{R4Red} and the terms, which arose due to Weyl rescaling \eqref{Weyl1} and \eqref{Weyl2}, and by using  the definition $ \mathcal{G}_{\mu\nu}  := R_{\mu \nu} - \frac{1}{4} g_{\mu \nu} R$, which is defined in close analogy to the Einstein tensor\footnote{ The Einstein tensor  is given by $G_{\mu \nu} =R_{\mu \nu} - \frac{1}{2} R g_{\mu \nu} $.}  one finds
\ba\label{R4Red3f}
S_{\text{kin}} =  \tfrac{1}{ (2 \pi)^4 \alpha'} \int_{M_4}  & \bls R     + \nabla_\mu  v^i  \nabla^\mu  v^j\tfrac{1}{ \cV}   \Big( \tfrac{1}{2}\cK_{ij} - \tfrac{1}{\cV}\cK_i\cK_j\Big) 
+ \tfrac{ \zeta(3) \; \alpha'}{4}  \, e^{- \tfrac{3}{2} \phi}\blr   \, \mathcal{G}_{\mu\nu} \mathcal{G}^{\mu \nu} \mathcal{Z}  -  \mathcal{G}_{\mu\nu}\,  \nabla^\nu  \nabla^\mu  v^i  \, \Big( \tfrac{2}{\cV} \cK_i  \mathcal{Z} \Big) & \nonumber \\
&\;\;\;\;\;+ \mathcal{G}_{\mu\nu}\, \nabla^\mu  v^i  \nabla^\nu  v^j \Big(   \mathcal{Z}_{ij}  - \tfrac{1}{\cV^2} \cK_i \cK_j   \mathcal{Z} \Big) - \tfrac{1}{2} \nabla_\mu\nabla^\mu  v^i \;\;  \nabla_\nu \nabla^\nu  v^j \Big(  \mathcal{Z}_{ij}  + \tfrac{1}{2 \cV^2}\cK_i \cK_j \mathcal{Z} \Big) & \nonumber \\
 &\;\;\;\;\;\;  +   \nabla_\mu\nabla_\nu  v^i \;\;  \nabla^\mu \nabla^\nu  v^j  \; \Big(  \mathcal{Z}_{ij}  + \tfrac{1}{ \cV^2}\ \cK_i \cK_j \mathcal{Z} \Big) \;\;  \brr \;\;  \brs \ast_4  1 \;\; .&
\ea
Where we have used the dimensionless quantities 
\ba\label{Zdef}
&\mathcal{Z} = \tfrac{1}{2 \pi \alpha'}\int_{Y_3} c_2 \wedge J  \; ,&
&\mathcal{Z}_{i}= \tfrac{1}{2 \pi \alpha'} \int_{Y_3} c_2 \wedge \omega_i \; ,&
&\mathcal{Z}_{ij} = -\tfrac{1}{4 \pi \alpha'} \int_{Y_3}\, Z_{m\bar m n \bar n} \o_i{}^{\bar m m} \o_j{}^{\bar n n} \ast  1 \;\; ,&
\ea
obeying the relations
\ba\label{relZs} 
 \mathcal{Z}_{i}= \mathcal{Z}_{ij}v^j =  \mathcal{Z}_{ji} v^j \;\;\; \text{and} \;\;\;    \mathcal{Z}=\mathcal{Z}_{i} v^i \;\; ,
\ea
which can be seen by using \eqref{Zrel}. Note that as expected $\frac{\d}{\d v^i} \cZ = \cZ_i \;\; \text{but} \;\;\; \frac{\d}{\d v^j}  \cZ_ i= 0$, thus  $\cZ_{ij}$ cannot be obtained easily by taking derivatives w.r.t.\,$\d v^i$. Let us stress that we have neglected $\alpha'$-corrections to the two-derivative part of this action \cite{Bonetti:2016dqh}, since those will not interfere with the four-derivative couplings. Furthermore, note that due to the uplift to finite fields $v^i$,  terms in \eqref{R4Red3f} may have a higher power in the fields $v^i$, in contrast to the quadratic dependence of the infinitesimal K\"{a}hler deformations.
Let us close this section by remaking that the higher-derivative effective action \eqref{R4Red3f} can be rewritten using field redefinitions involving higher-derivative pieces themselves.  Thus the  given presentation is a particular choice, which results naturally after dimensional reduction. However, one may perform field redefinitions as e.g.
\beq
g_{\mu \nu} \to g_{\mu \nu} + a R_{\mu \nu} + b R g_{\mu \nu}  \;\;  a,b \in \mathbb{R} \;\; .
\eeq
One concludes that the higher-derivative couplings  in \eqref{R4Red3f} are presented in one particular frame of the fields $g_{\mu \nu}$ and $v^i$. A more sophisticated analysis of the supersymmetric completion at the four-derivative level would be required to select a canonical frame.

 \section{The 4d, $\cN =1$ action}\label{4dN1}

In this section we perform the orientifold projection on the effective action \eqref{R4Red3f}, which amounts to adding $O3/O7$ planes  to the Calabi-Yau background \cite{Giddings:2001yu,Acharya:2002ag,Brunner:2003zm,Brunner:2004zd,Dabholkar:1996pc}. For consistency we are required to  also consider $D3/D7$ branes in this setup. However, we will not discuss any $\alpha'^3$-corrections arising from these sources, but let us emphasize that a complete treatment would require such a refined analysis. Already at the classical level these would source a warp-factor and background fluxes, which we chose not to account for.

In \ref{orientifold} we review the well known properties of the orientifold projection on Calabi-Yau threefolds \cite{Grimm:2004uq,Andrianopoli:2001zh}, and apply it to the four-derivative effective action derived in the previous section.
We then proceed in  \ref{LinMul} by expressing the truncated spectrum in terms of the real scalar fields of the linear multiplet of $4d$,  $\cN =1$ supergravity. 
 
\subsection{ Orientifold projection}\label{orientifold}

In the following we consider $O3/O7$ planes in the Calabi-Yau threefold background, known as Calabi-Yau orientifold, denoted in the following as $X$. The presence of orientifold planes truncates the effective theory from $\cN=2$ to $\cN=1$ supersymmetry.
Orientifold planes manifest themselves as an isometric, holomorphic involution $\sigma: X\to X$,  thus $\sigma^2= id$ and $ \sigma^\ast g = g$ on the internal Calabi-Yau space with metric $g$, such that
 \beq\label{oProj}
 \sigma^\ast J =J \;\;.
  \eeq
 Moreover, the presence of $O3/O7$ planes results in $\sigma^\ast \Omega = - \Omega$, where $\Omega$ is the holomorphic $(3,0)$-form. Furthermore,  considering the action of $\Omega_p (-1)^{F_L}$ on the space-time fields, where $\Omega_p$ is the world-sheet parity and $F_L$ the space-time fermion
 number of the left moving sector, one finds that
 \beq\label{oProj2}
\Omega_p (-1)^{F_L} \phi =\phi \;\;\;  \text{and}  \;\;\; \Omega_p (-1)^{F_L} g=g \;\; .
 \eeq
 The cohomology groups $H^{p,q}$ naturally decompose in odd and even eigenspaces under the action of $\sigma^\ast$ as $H^{p,q} =  H^{p,q}_{+} \oplus H^{p,q}_{-} $. Since the K\"{a}hler form is invariant under the orientifold projection \eqref{oProj}, only the K\"{a}hler deformations related to the even eigenspace $H^{1,1}_+$ remain in the spectrum, such that $J = v^a \o_a ,\;\;a =1,\dots,h^{1,1}_+$.
 
Subjected to the orientifold projection the reduction result \eqref{R4Red3f}  has to be modified accordingly and one straightforwardly arrives at
 \ba\label{R4Red4}
S_{\text{kin}} = \tfrac{1}{ (2 \pi)^4 \alpha'} \int_{M_4}  & \bls R     + \nabla_\mu  v^i  \nabla^\mu  v^b\tfrac{1}{ \cV}   \Big( \tfrac{1}{2}\cK_{ab} - \tfrac{1}{\cV}\cK_a\cK_b\Big) 
+\tfrac{ \zeta(3) \, \alpha'}{4}  e^{- \tfrac{3}{2} \phi}\blr   \, \mathcal{G}_{\mu\nu} \mathcal{G}^{\mu \nu} \mathcal{Z}  -  \mathcal{G}_{\mu\nu}\,  \nabla^\nu  \nabla^\mu  v^a  \, \Big( \tfrac{2}{\cV} \cK_a  \mathcal{Z} \Big) &\nonumber  \\
&\;\;\;\;\;\; + \mathcal{G}_{\mu\nu}\, \nabla^\mu  v^a  \nabla^\nu  v^b \Big(   \mathcal{Z}_{ab} - \tfrac{1}{\cV^2} \cK_a \cK_b   \mathcal{Z} \Big)  - \tfrac{1}{2} \nabla_\mu\nabla^\mu  v^a \;\;  \nabla_\nu \nabla^\nu  v^b \Big(  \mathcal{Z}_{ab} + \tfrac{1}{2 \cV^2}\cK_a \cK_b \mathcal{Z} \Big)  &\nonumber \\ 
&\;\;\;\;\;\; +   \nabla_\mu\nabla_\nu  v^a \;\;  \nabla^\mu \nabla^\nu  v^b  \; \Big(\mathcal{Z}_{ab} +  \tfrac{1}{ \cV^2} \cK_a \cK_b \mathcal{Z}\Big) \;\; \brr \;\;\brs \ast_4  1 \;\; .& 
\ea
Where we have used the properties of the orientifold projection to conclude that
\ba\label{ZorProj}
&\mathcal{ Z} =  \tfrac{1}{2 \pi \alpha'} \int_{Y_3} c_2 \wedge J =  \tfrac{1}{2 \pi \alpha'} \int_{X} c_2 \wedge J \;\; , \quad \quad \;\;  \mathcal{Z}_{a}=  \tfrac{1}{2 \pi \alpha'} \int_{Y_3} c_2 \wedge \omega_a =\tfrac{1}{2 \pi \alpha'} \int_{X} c_2 \wedge \omega_a  & \\  \nonumber
&\mathcal{Z}_{ab } = - \tfrac{1}{4 \pi \alpha'} \int_{Y_3}\, Z_{m\bar m n \bar n} \o_a{}^{\bar m m} \o_b{}^{\bar n n} \ast  1 = -\tfrac{1}{4 \pi \alpha'}\int_{X}\, Z_{m\bar m n \bar n} \o_a{}^{\bar m m} \o_b{}^{\bar n n} \ast  1 \;\; , &
 \ea
obeying the analogous relations to \eqref{relZs} given by
\ba
 \mathcal{Z}_{a}= \mathcal{Z}_{ab}v^b =  \mathcal{Z}_{ba} v^b \;\;\; \text{and} \;\;\;    \mathcal{Z}=\mathcal{Z}_{a} v^a \;\; .
\ea

\subsection{4d, $\cN=1$  linear multpilets}\label{LinMul}

The canonical form of the $4d, \,\cN=1$ action for the real scalars  $L^a$  in the  linear multiplets  takes the form
\beq
S =   \tfrac{1}{ (2 \pi)^4 \alpha'}  \int_{M_4}   R \ast 1 \;  + \;  \tfrac{1}{2}  G_{ab} \nabla_\mu  L^a  \nabla^\mu L^b  \ast 1\; ,
\eeq
with the couplings $ G_{ab} $, which can be inferred from a kinematic potential $\tilde K $ as $G_{ab}  = \frac{\d}{\d L^a} \frac{\d}{\d L^b} \tilde K$.
The identification of the K\"{a}hler moduli fields $v^a$ with the  real scalars  in the linear multiplet of the $4d$, $ \cN=1$ supergravity theory at leading order in $\alpha'$ is given by 
 \beq \label{Lredef}
 L^a =  \frac{v^a}{\cV} \;\; .
 \eeq
Eventual $\alpha'$-modifications of \eqref{Lredef} due to the two-derivative analysis at this order in $\alpha'$ \cite{Bonetti:2016dqh} do not alter the four-derivative couplings at the relevant order in $\alpha'$, thus it suffices to express the action in terms of \eqref{Lredef}.
 To determine all the relevant four-derivative couplings of $L^i$ one requires knowledge of the couplings cubic and quartic in the infinitesimal fluctuations $\d v^i $ arising from the $\cR^4$-sector. This is however, beyond the study of this work and we have  thus omitted such terms also arising due to the Weyl rescaling  in \eqref{Weyln}. However, one may show that one can express the couplings $  T_{\mu \nu }\,  \nabla^\nu  \nabla^\mu  v^a $ and $T_{\mu \nu }\, \nabla^\mu  v^a  \nabla^\nu  v^b $ in terms of the fields $L^a$  in the linear multiplets without making use of information of the neglected sector. This does not apply to the $\nabla_\mu\nabla^\mu  v^a \;\;  \nabla_\nu \nabla^\nu  v^b$  and $\nabla_\mu\nabla^\nu v^a \;\;  \nabla_\nu \nabla^\mu  v^b $ terms where the knowledge of the other four-derivative couplings is crucial.  Hence  we will not consider  the latter in the following.
Expressing \eqref{R4Red4} in terms of the linear multiplets one finds
\ba\label{R4Red5}
S_{\text{kin}} =   \tfrac{1}{ (2 \pi)^4 \alpha'}  \int_{M_4}  \bls & R \;     + \tfrac{1}{2} \nabla_\mu  L^a  \nabla^\mu L^b  \;\; \cV \Big(   \cK_{ab} -  \tfrac{1}{\cV}\cK_a\cK_b \Big)+   \tfrac{ \zeta(3) \alpha'}{4} e^{- \frac{3}{2} \phi}  \blr  \ \, \mathcal{G}_{\mu\nu} \mathcal{G}^{\mu \nu} \mathcal{Z} +  \mathcal{G}_{\mu\nu}\,  \nabla^\nu  \nabla^\mu   L^a  \, \cK_a   \mathcal{Z}  \nonumber
&  \\ 
& \;\;\; \;\;+  \mathcal{G}_{\mu\nu}\, \nabla^\mu  L^a  \nabla^\nu  L^b \Big(  \cV^2  \mathcal{Z}_{ab} + \tfrac{5}{2} \cK_a \cK_b \mathcal{Z}  - 3 \cV  \cK_{ab}   \cZ-  \cV\cK_a \mathcal{Z}_b  \Big) \;\; \brr \;\; \brs \ast1  \ . &  
\ea
 Classically one then encounters the K\"{a}hler metric on the moduli space to be given by
 \beq
G_{ab} =  \cV \int_{X} \o_a \wedge \ast \o_b =  \cV \Big(   \cK_{ab} -  \tfrac{1}{\cV}\cK_a\cK_b \Big)\;\; ,
\eeq
arising from the kinematic potential  $\tilde K = -2 \log \cV = \log \cK_{ijk}L^iL^jL^k$. 
The resulting novel couplings at order $\alpha'^3$, couple derivatives of the real scalars $L^a$ to the tensor $\mathcal{G}_{\mu\nu}$, which is composed  out of the Ricci tensor and the Ricci scalar. The higher-derivative coupling $\mathcal{G}_{\mu\nu} \mathcal{G}^{\mu \nu} \mathcal{Z} $ has been analyzed in \cite{Alvarez-Gaume:2015rwa}, and leads to a propagating massive spin 2 ghost mode. However, let us note  that the appearance of ghost modes in effective field theories is not an immediate issue since it is related to the truncation of the ghost-free infinite series resulting from string theory.

Let us  next comment on the term  $\mathcal{G}_{\mu\nu}\, \nabla^\mu  L^a  \nabla^\nu  L^b $. 
Firstly, note that this higher-derivative coupling does not correct the propagator of $L^a$, since it vanishes in the Minkowski background. Thus it does  not give rise to  any ghost modes for $L^a$. 
 The analogous  case of the Einstein-tensor coupled to a scalar field  is well studied and relevant in the context of inflation. It was observed that such a coupling of a scalar field to curvature terms favors slow roll inflation, in other words rather steep potentials can exhibit the feature of slow roll.
It is expected that this coupling \eqref{R4Red5} could be used to implement these scenarios in the context of K\"{a}hler moduli inflation. It is an old approach in the context of string theory to drive slow roll inflation by a K\"{a}hler modulus \cite{PhysRevD.34.3069,PhysRevD.52.3548,Conlon:2005jm,Cicoli:2008gp,Burgess:2016owb}. 
It would be interesting to analyze the consequences of the derived novel couplings to such inflationary models and their relevance due to their $\alpha'^3$-suppression \cite{Cicoli:2015wja,Broy:2015zba,Cicoli:2016chb}.
Finally, let us discuss the coupling $ \mathcal{G}_{\mu\nu}\,  \nabla^\nu  \nabla^\mu   L^a $. As in the above case it does not correct the propagator of $L^a$. In contrast to the previous case these couplings are poorly studied in inflation literature and hence their embedding in string inflation models is desirable. 
In both cases coefficients dependent on topological quantities $\cZ , \cZ_a$, see   \eqref{ZorProj}, of the internal Calabi-Yau orientifold  and are trivially related to the analog quantities \eqref{Zdef} of the Calabi-Yau threefold,
 and are thus  computable in the context of algebraic geometry.
 However, the semi-topological coupling $\cZ_{ab}$ requires the knowledge of the Calabi-Yau metric and although derivable in principle it is beyond the capability of current available techniques.

\section{Conclusions}

Considering purely gravitational $\cR^4$-corrections at order $\alpha'^3$ to the leading order IIB supergravity action in ten dimensions, we performed a dimensional reduction to four dimensions on a Calabi-Yau threefold. Analyzing the reduction result at four-derivative level and quadratic in the infinitesimal  K\"{a}hler deformations
we derived novel  couplings  of the K\"{a}hler moduli fields and gravity. We argued that these are complete in a sense that the couplings cannot be altered by other sectors of the IIB action at order $\alpha'^3$, or by modifications of the background. 
We then performed the orientifold projection to derive a minimal supergravity theory in four dimensions. Let us stress that for a complete analysis one needs to derive the reduction result up to quartic order in the infinitesimal K\"{a}hler deformations. Only then one is able to draw definite conclusions for all of the resulting four-derivative couplings involving the K\"{a}hler moduli fields and gravity. This is an interesting question to be answered and the obvious next step in this research program. Let us conclude by emphasizing that a detailed analysis of the novel couplings in the context of K\"{a}hler moduli inflation in IIB orientifold setups is desirable.

%%%%%%%%%%%%%%%%%%%%%%%%%%%%%%%%%%%%%%%%%%
\vspace*{.5cm}
\noindent
\subsection*{Acknowledgments}
%%%%%%%%%%%%%%%%%%%%%%%%%%%%%%%%%%%%%%%%%%%%%%%%

I would like to thank Federico Bonetti, Thomas Grimm, Dieter Luest,  Taizan Watari  and  Itamar Yakov for helpful discussions
and comments. In particular, I would like to express my thankfulness to the theoretical high-energy physics groups of the Walter Burke institue at Caltech, the center for the fundamental laws of nature in Harvard, and the Max-Planck institute for physics in Munich, for their hospitality  during my visits.   This work was supported by the Grant-in-Aid for Scientific Research on Innovative Areas 2303, MEXT, and the WPI program of Japan.

%%%%%%%%%%%%%%%%%%%%%%%%%%%%%%%%%%%%%%%%%%%%%%%%
\begin{appendix}
\vspace{2cm} 
\noindent {\bf \LARGE Appendix}
\addcontentsline{toc}{section}{Appendix}
%%%%%%%%%%%%%%%%%%%%%%%%%%%%%%%%%%%%%%%%%%%%%%%%
\section{\bf Conventions, definitions, and identities} \label{Conv_appendix}

In this work we denote the ten-dimensional space indices by capital Latin letters $M,N = 0,\ldots,9$,
the external  ones by  $\mu,\nu = 0,1,2,3$, and the internal complex ones by $m,n,p=1,2,3$ and $\bar m, \bar n,\bar p =1, 2,3$. The metric signature of the ten-dimensional space  is $(-,+,\dots,+)$.
Furthermore, the convention for the totally 
anti-symmetric tensor in Lorentzian space in an orthonormal frame is $\epsilon_{012...9} = \epsilon_{012}=+1$. 
The epsilon tensor in $d$ dimensions then satisfies
\ba
\epsilon^{R_1\cdots R_p N_{1 }\ldots N_{d-p}}\epsilon_{R_1 \ldots R_p M_{1} \ldots M_{d-p}} &= (-1)^s (d-p)! p! 
\delta^{N_{1}}{}_{[M_{1}} \ldots \delta^{N_{d-p}}{}_{M_{d-p}]} \,, 
\ea
where  $s=0$ if the metric has Riemannian signature and $s=1$ for a Lorentzian metric.
We adopt the following conventions for the Christoffel symbols and Riemann tensor 
\ba
\G^R{}_{M N} & = \fr12 g^{RS} ( \pa_{M} g_{N S} + \pa_N g_{M S} - \pa_S g_{M N}  ) \, , &
R_{M N} & = R^R{}_{M R N} \, , \nonumber\\
R^{M}{}_{N R S} &= \pa_R \G^M{}_{S N}  - \pa_{S} \G^M{}_{R N} + \G^M{}_{R  T} \G^T{}_{S N} - \G^M{}_{ST} \G^T{}_{R N} \,, &
R & = R_{M N} g^{M N} \, , 
\ea
with equivalent definitions on the internal and external spaces. Written in components, the first and second  Bianchi identity are
\bea\label{Bainchiid}
{R^O}_{PMN} + {R^O}_{MNP}+{R^O}_{NPM} & = & 0 \nonumber\\
(\nabla_L R)^O{}_{PMN} + (\nabla_M R)^O{}_{PNL} + (\nabla_N R)^O{}_{PLM} & = & 0 \;\;\; .
\eea

Let us specify in more detail our conventions regarding complex coordinates
in the internal space.
For a 
complex Hermitian manifold $M$ with complex dimension $n$
the complex coordinates $z^1 , \dots, z^n$ and 
the underlying real coordinates $\xi^1, \dots , \xi^{2n}$ are related by
\begin{equation}
( z^1,...,z^n ) = \left(  \frac{1}{\sqrt{2}}(\xi^1 + i \xi^2), \dots ,  \frac{1}{\sqrt{2}}(\xi^{2n-1} + i \xi^{2n}) \right) \,.
\end{equation}
Using these conventions one finds
\begin{equation}
\sqrt{g}  d\xi^1 \wedge ... \wedge d\xi^{2n} = \sqrt{g} (-1)^{\frac{(n-1)n}{2}} i^n  dz^1\wedge...\wedge dz^n 
\wedge d\bar z^1 \wedge...\wedge d\bar z^n = \frac{1}{n!} J^n \,,
\end{equation}
with $g$ the determinant of the metric in real coordinates and  $\sqrt{\det g_{m n}} = \det g_{m \bar n} $. The K\"{a}hler form is given by
\begin{equation}
\label{eq:Kform}
J = i g_{m\bar{n} } dz^m \wedge d\bar z^{\bar{n} } \, .
\end{equation}
Let $\omega_{p,q}$ be a $(p,q)$-form, then its Hodge dual is the $(n-q,n-p)$ form
\begin{align} \label{eq:pgform}
\ast \omega_{p,q} & = \frac{ (-1)^{\frac{n(n-1) }{2}  } \, i^n \, (-1)^{pn}}  {p!q!(n-p)!(n-q)!}   
\omega_{m_1 \dots m_p \bar{n} _1 \dots \bar{n} _q} 
\epsilon^{m_1 \dots m_p}_{\phantom{m_1 \dots m_p} \bar r_1 \dots \bar r_{n-p}} \nonumber \\
& \quad \times \epsilon^{\bar{n} _1 \dots \bar{n} _q}_{\phantom{\bar \beta_1 \dots \bar{n} _q}  s_1 \dots  s_{n-q}} 
dz^{ s_1}\wedge \dots \wedge dz^{ s_{n-q}} \wedge d \bar z^{\bar r_1} \wedge \dots \wedge d \bar z^{\bar  r^{n-p}}.
\end{align}

Finally, let us record our conventions regarding Chern forms.
To begin with, 
 we define the curvature two-form for Hermitian manifolds to be
  \begin{equation}\label{curvtwo}
 {\cR^m}_n  =  {{R^m}_n }_{ r \bar s} dz^ r \wedge d\bar{z}^\bar{s}\;\;,
  \end{equation}
and we set
 \bea \label{defR3}
 \Tr{\cR}\;\;&  =& {{R^ m }_ m }_{ r \bar{s}}dz^ r \wedge d\bar{z}^{\bar{s}} \;,\nonumber \\
 \Tr{\cR^2} &= & {{R^{ m }}_{n }}_{ r \bar{s}} {{R^{n }}_{ m }}_{ r_1 \bar{s}_1}dz^{ r}
 \wedge d\bar{z}^{\bar{s}}\wedge dz^{ r_1} \wedge d\bar{z}^{\bar{s}_1} \;,\nonumber  \\
 \Tr{\cR^3} &=& {{R^{ m }}_{n }}_{ r \bar{s}}  R^{n }{}_{n _1  r_1 \bar{s}_1}
 {{R^{n _1}}_{ m }}_{ r_2 \bar{s}_2}dz^{ r} \wedge d\bar{z}^{\bar{s}}\wedge dz^{ r_1} \wedge d\bar{z}^{\bar{s}_1}\wedge dz^{ r_2} \wedge d\bar{z}^{\bar{s}_2} \; .
 \eea
 The Chern forms can then  be expressed in terms of the curvature two-form as
\begin{align}  \label{Chernclasses}
 c_0 &= 1\nonumber \;, \\
 c_1 &= \frac{1}{2\pi} i \Tr{ \mathcal{R}} \nonumber\;, \\
 c_2 &= \frac{1}{(2\pi)^2} \frac{1}{2}\left( \Tr{\cR^2} -  (\Tr{\cR})^2 \right)\;,  \nonumber\\
 c_3 &=   \frac{1}{3}c_1c_2 + \frac{1}{(2\pi)^2} \frac{1}{3} c_1 \wedge \Tr \cR^2 - 
 \frac{1}{(2\pi)^3}\frac{i}{3} \Tr \cR^3 
\end{align}
The Chern forms of an $n$-dimensional Calabi-Yau manifold $Y_n$ reduce to
\beq\label{chern34}
c_2 (Y_{n \geq 2}) = \frac{1}{(2\pi)^2} \frac{1}{2} \Tr{\cR^2}  \;\; \text{and} \;\;c_3 (Y_{n \geq 3}) =  - \frac{1}{(2\pi)^3} \frac{i}{3}  \Tr{\cR^3} 
\eeq
The six dimensional Euler-density is given by 
\beq \label{Euler}
Q =  -\tfrac {1}{3} \left(  R^\tbzero_{m_1}{}^{m_2}{}_{n_1}{}^{n_2} 
R^\tbzero_{m_2}{}^{m_1} {}_{n_2} {}^{n_3}
R^\tbzero_{n_2} {}^{n_1} {}_{n_3} {}^{n_2} + R^\tbzero_{m_1}{}^{m_2}{}_{n_1}{}^{n_2} 
R^\tbzero_{m_2}{}^{m_3} {}_{n_2} {}^{n_3}
R^\tbzero_{m_3} {}^{m_1} {}_{n_3} {}^{n_1}  \right) \ .
\eeq
It satisfies
\beq \label{Q_integral}
Q = (2\pi)^3  \, *_6 c_3 \ , \qquad
\int_{Y_3} Q *_6 1 = (2\pi)^3 \chi \ ,
\eeq
where $\chi$ is the Euler-Characteristic of the internal  Calabi-Yau manifold.
Let us next define the intersection numbers\ba\label{IN1}
\cK_{i j k}  &=  \tfrac{1}{(2 \pi \alpha')^3}\int_{Y_3} \o_i \we   \o_j \we   \o_k \, , \quad \quad \;\;\; \quad \quad \quad 
\cK_{i j }   =  \tfrac{1}{(2 \pi \alpha')^3}\int_{Y_3} \o_i \we   \o_j \we   J =  \cK_{i j k } v^k \,,  \nonumber  \\
\cK_{i } \;\; & =   \tfrac{1}{2 ( 2 \pi \alpha')^3}\int_{Y_3} \o_i \we  J\we   J  = \fr12 \cK_{i j k } v^j v^k \, , \quad 
\cV \;= \tfrac{1}{3! ( 2 \pi \alpha')^3}\int_{Y_3}  J \we  J\we  J  = \fr1{3!} \cK_{i j k } v^i v^j v^k  \, ,
\ea 
where $\{ \o_i \}$ are harmonic  (1,1) -forms w.r.t.~to the Calabi- Yau metric $g_{m \bar n}$.
Let us state the useful identities
\ba \label{IN2}
\omega_{im}{}^m & = i \frac{\cK_i}{\cV} \ , &
\omega_{im \bar n} \omega_j{}^{\bar nm} \, *_61 & = 
\o_i \wedge \o_j \wedge J- \frac{1}{\cV{}^2} \cK_i \cK_j \,
 *_61 \ .
\ea

We present the formulae for a Weyl rescaling  $g_{\mu \nu} \to \Omega g_{\mu \nu}$ of the four-derivative terms, $R_{\mu \nu} R^{\mu \nu} ,\,  R^2$. These expressions can be derived straight forwardly, and are given by
\bea\label{Weyl1}
R^2 \stackrel{\text{Weyl}}{\rightarrow} & \frac{1}{\Omega^2} R^2  \,\, - \,\,    6 \; R \frac{1}{\Omega^3} ( \nabla_{\mu} \nabla^\mu \Omega ) 
 \,\, + \,\,  3\frac{1}{\Omega^4}  R  \,\, ( \nabla_{\mu} \Omega )(\nabla^{\mu}\Omega)
 \,\, +   \,\,  9  \frac{1}{\Omega^5}  (\nabla_{\mu} \nabla^\mu \Omega ) (  \nabla_{\nu} \nabla^\nu \Omega ) &\nonumber \\
& \,\,  -  \,\,  9  \frac{1}{\Omega^5} ( \nabla_{\mu} \Omega)( \nabla^{\mu}\Omega ) (  \nabla_{\nu} \nabla^\nu \Omega )
  \,\, +  \,\, 9   \frac{1}{4 \Omega^6}( \nabla_{\mu} \Omega)( \nabla^{\mu}\Omega)  ( \nabla_{\nu} \Omega)( \nabla^{\nu}\Omega ) \;\; ,& \;\; 
\eea
and
\bea\label{Weyl2}
R_{\mu \nu} R^{\mu \nu}  \stackrel{\text{Weyl}}{\rightarrow} & \frac{1}{\Omega^2} R_{\mu \nu} R^{\mu \nu}  \,\, - \,\,     \; R \frac{1}{\Omega^3} ( \nabla_{\mu} \nabla^\mu \Omega ) 
 \,\, + \,\,  3 \frac{1}{\Omega^4}  R_{\mu \nu}  \,\,  \nabla
^{\mu} \Omega \nabla^{\nu}\Omega - \,\,  2  \frac{1}{\Omega^4}  R_{\mu \nu}  \,\,  \nabla
^{\mu} \nabla^{\nu}\Omega &\nonumber \\
& \,\, +   \,\,  2  \frac{1}{\Omega^5}  (\nabla_{\mu} \nabla^\mu \Omega ) (  \nabla_{\nu} \nabla^\nu \Omega ) 
  \,\, +   \,\,    \frac{1}{\Omega^5}  (\nabla^{\mu} \nabla^\nu \Omega ) (  \nabla_{\mu} \nabla_\nu \Omega ) 
 \,\,  -  \,\,  \frac{3}{2}  \frac{1}{\Omega^5} ( \nabla_{\mu}  \nabla^{\mu}\Omega ) (  \nabla_{\nu}\Omega )( \nabla^\nu \Omega )& \nonumber \\
 &
  \,\,  -  \,\,   3  \frac{1}{\Omega^5} ( \nabla^{\mu}  \nabla^{\nu}\Omega ) (  \nabla_{\nu}\Omega )( \nabla_\mu \Omega )
  \,\, +  \,\, 9   \frac{1}{4 \Omega^6}( \nabla_{\mu}  \Omega ) (\nabla^{\mu}\Omega)  ( \nabla_{\nu} \Omega)( \nabla^{\nu}\Omega )\;\; .& 
\eea

%%%%%%%%%%%%%%%%%%%%%%%%%%%%%%%%%%%%%%%%%%%%%%%%
\end{appendix}
%%%%%%%%%%%%%%%%%%%%%%%%%%%%%%%%%%%%%%%%%%%%%%%%

\bibliography{Matthias}
\bibliographystyle{utcaps}

%%%%%%%%%%%%%%%%%%%%%%%%%%%%%%%%%%%%%%%%%%%%%%%%
\end{document}